\documentclass[twocolumn,showpacs,preprintnumbers,amsmath,amssymb]{revtex4-1}

\usepackage{graphicx}% Include figure files
\usepackage{dcolumn}% Align table columns on decimal point
\usepackage{bm}% bold math
\usepackage[T2A]{fontenc}
\usepackage[cp1251]{inputenc}
\usepackage[dvips]{color}

\usepackage[normalem]{ulem} % to cross out sentences

\begin{document}

\title{Feedback spectroscopy of atomic resonances}

\author{V. I.~Yudin$^{1,2,3,4}$, A. V.~Taichenachev$^{1,2,4}$, D. I.~Sevostianov$^{5,6}$,  V.~L.~Velichansky$^{4,5,6}$, V.~V.~Vasiliev$^{6}$,  A. A.~Zibrov$^{4,7}$, A. S.~Zibrov$^{2,6,8}$, S. A.~Zibrov$^{4,6}$ \\
 $^{1}${\em Institute of Laser Physics, Siberian Branch of RAS,  Novosibirsk, 630090, Russia}\\
 $^{2}${\em Novosibirsk State University, Novosibirsk, 630090, Russia}\\
 $^{3}${\em Novosibirsk State Technical University, Novosibirsk, 630092, Russia}\\
 $^{4}${\em Russian Quantum Center, Skolkovo, Moscow Reg., 143025, Russia}\\
 $^{5}${\em National Research Nuclear University (MEPhI), Moscow, 115409, Russia }\\
 $^{6}${\em Lebedev Physical Institute, RAS, Moscow, 117924, Russia}\\
 $^{7}${\em Center for Astrophysics, Harvard University, Cambridge, MA, 02138, USA}\\
 $^{8}${\em Physics Department, Harvard University, Cambridge, MA, 02138, USA}\\
}
\date{\today}

\begin{abstract}

We propose a non-standard spectroscopic technique that uses a feedback control of the input probe field parameters to significantly increase the contrast and quality factor of the atomic resonances. In particular, to apply this technique for the dark resonances we sustain the fluorescence intensity at a fixed constant level while taking the spectra process. Our method, unlike the conventional spectroscopy, does not require an optically dense medium. Theoretical analysis has been  experimentally confirmed in spectroscopy of atomic rubidium vapor in which a considerable increase (one-two order) of the resonance amplitude and a 3-fold decrease of the width have been observed in optically thin medium. As a result, the quality factor of the dark resonance is increased by two orders of magnitude and its contrast reaches a record level of 260$\%$. Different schemes, including magneto-optical Hanle spectroscopy and Doppler-free spectroscopy have also showed a performance enhancement by using the proposed technique.

\end{abstract}


\maketitle

Over time a spectroscopy  has  established a certain requirement to take a medium's response as a function of frequency of the probing field while the rest of the input parameters (intensity, polarization, spatial distribution, etc.) are kept at a constant level. Usually in this case the registered signals are either the absorption or the fluorescence spectra. Instead, we suggest to hold the medium response at a constant level (during the frequency scanning) by manipulation of the input probing field via electronic feedback control. In this case the changes of the governed input parameters imitate the medium's spectrum. To test our conception we have applied it to a well-known phenomenon - the coherent population trapping (CPT) \cite{Alzetta,Arimondo,Harris,Fleisch}.

The main feature of the CPT consists in the existence of so-called dark state $|dark\rangle$, which is a coherent superposition state and nullifies an atomic-light interaction operator $\hat{V}_{}$: $\hat{V}_{}|dark\rangle=0$. In the dark state the atoms neither absorb nor emit a light. For the modern laser metrology the importance of CPT lies in the development of miniature (including chip-scale) atomic clocks \cite{Vanier-Review1,Vanier-Review2,Knappe,Shah,Novikova2,Symmetricom} and magnetometers \cite{Scully,Novikova,Stahler2,Budker2,Bud,Yudin}. These devices are based on a two-photon resonance (for rubidium or cesium atoms, above all) formed in a bichromatic laser field, in which the frequency difference of the spectral components ($\omega_1-\omega_2$) is varied near the hyper-fine splitting $\Delta$. For such spectroscopy
the existence of a pure coherent state $|dark\rangle$, which is
sensitive to the two-photon detuning $\delta_\text{R}$=($\omega_1-\omega_2-\Delta$),
leads to a significant increase of the contrast and quality
factor (amplitude-to-width ratio) of the CPT resonance in combination
with a decrease of the its light shift. Namely it explains why for alkali atoms the D$_1$ line is
much preferable in comparison with the D$_2$ line, for which the pure
dark state is absent in the cases of
Doppler and/or collisional broadening of optical line \cite{Stahler,Y}. Pursuing a higher resonance contrast, the new polarization schemes have been implemented \cite{Y,Jau,Taich2004,Zanon,lin_lin JETP,Serezha1,Zibrov,Shah2}, where is
no ``trap'' Zeeman state (with the maximal projection of the angular
moment $m=F$), which is insensitive to the two-photon detuning.
The Ramsey effect also narrows the CPT resonance and
gives some increase of its quality \cite{Matsko,Zanon,Ron,Breschi,Grujic}.

However to date, within the bounds of traditional spectroscopy the ways to improve the CPT resonances are likely exhausted.
A further improvement of the quality factor can be achieved by increasing the number of atoms interacting with light, that contradicts to the goals of miniaturization and power consumption of the CPT-based atomic clocks and magnetometers. Moreover, although the high atomic density gives a some increase of the resonance contrast, in an optically thick medium the nonlinear effects distort the resonance line-shape \cite{Lukin,Vanier-Review1} and, consequently, reduce the metrological characteristics of the resonance.

As we mentioned above, in this paper we propose an alternative spectroscopic technique, which was tested in detail for the CPT resonance. In particular, while the conventional spectroscopy uses the fixed input light intensity to record the spectra, we use an electronic feedback control \textit{to change} the input intensity in a way such that the level of the spontaneous fluorescence is constant during the frequency scanning. This results in a radical increase of the CPT resonance contrast together with narrowing of its width. Our method does not require a high optical density and is quite effective in small atomic cells or at lower temperatures, thus opening new opportunities for the laser spectroscopy and metrology.

\begin{figure}[t]
\centerline{\scalebox{0.5}{\includegraphics{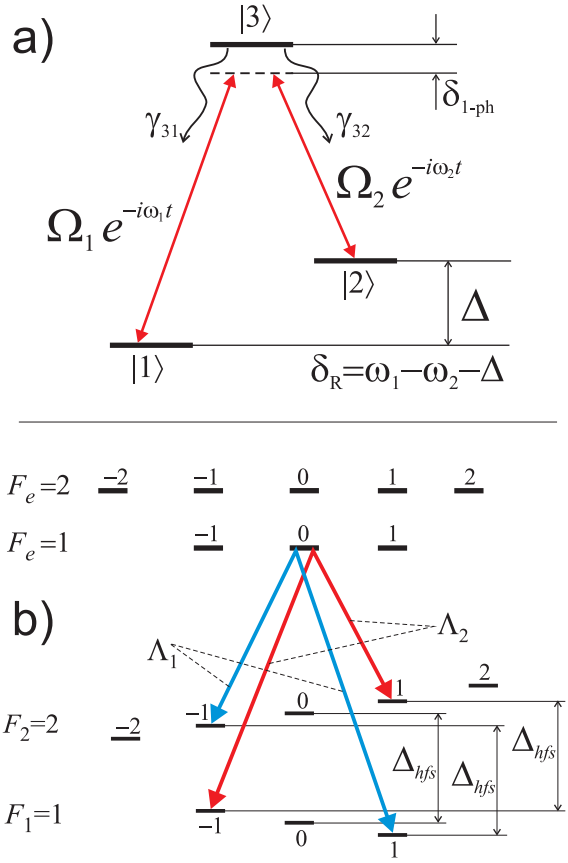}}}\caption{Shemes for creating of the two-photon dark resonance:\\
a) simplified three-level $\Lambda$ system;\\
b) two $\Lambda$ systems at the $D_1$ line of $^{87}$Rb for $Lin||Lin$ field configuration, which was used in experiment.
Here we do not show Zeeman shifts for upper hyperfine levels with $F_e$=1,2.}\label{Lambda}
\end{figure}

\section{Theory}

To explain the basic idea of our method let us consider the general case, when an atomic medium is illuminated by the light with frequencies $\{\omega_j\}$. Each component $\omega_j$ has a number of the different input parameters (power, polarization, phase, spatial distribution, etc.) to which we assign the unified symbol $Q_j$. All possible responses of the medium can be formally represented as the functions $S_{a}(\omega_1,...\omega_j...;Q_1,...Q_j...)$, where the index $a$ labels a type of the medium's response (such as absorption, spontaneous emission, output intensity and phase, polarization, refraction index and so on). In these terms, the conventional spectroscopy studies the frequency dependencies $S_{a}(\omega_1,...\omega_j...;Q_1,...Q_j...)$ at a set of constant input parameters $\{Q_j\}=const$.

In contrast, we propose another scenario, when at least one of medium's responses $S_{b}$ is fixed at a constant level (during frequency scanning) by applying feedback to the manipulated input parameters $Q_j$:
\begin{equation}\label{condition}
	S_b(\omega_1,...\omega_j...;Q_1,...Q_j...)=const,  ~\{Q_j\}\neq const\,.
\end{equation}
In this case the eq.~(\ref{condition}) defines the frequency dependencies of the input parameters $Q_j(\omega_1, ... \omega_j ...)$, which imitate an atomic spectrum. Besides $Q_j(\omega_1, ... \omega_j ...)$, in such feedback spectroscopy we can also detect the frequency dependencies of other medium's responses $S_{a'}$, where $a'\neq b$ in eq.~(\ref{condition}).

To our knowledge an approach similar to ours had been used only once before. In a saturation interference experiment \cite{Schawlow} the light's frequency and phase had been controlled simultaneously. As a result, the spectral resolution had been improved by a factor of 1.5 compared to the traditional saturation spectroscopy.

Let us apply our method to the dark resonance, which is formed by the bichromatic traveling wave with close frequencies $\omega_1\simeq\omega_2$:
\begin{equation}\label{E}
E(t)=f(x,y)(E_1 e^{-i\omega_1 t}+E_2 e^{-i\omega_2 t})e^{ikz}+c.c.\,,
\end{equation}
where $E_{j}$ is the amplitude of the $j$-th frequency component ($j$=1,2), $k$ is the wave number, and the function $f(x,y)$ describes the transverse field distribution in the light beam (step-like, Gaussian, and so on).
We consider the standard three-level $\Lambda$ scheme (see. Fig.~\ref{Lambda}a) with two optical transitions $|1\rangle\leftrightarrow |3\rangle$, $|2\rangle\leftrightarrow |3\rangle$, which are resonant to the $\omega_1$ and $\omega_2$, respectively.  The Rabi frequencies are defined as $\Omega_1=d_{31}E_1$ and $\Omega_2=d_{32}E_2$, where  $d_{31}=\langle 3|\hat{d}|1\rangle$ and $d_{32}=\langle 3|\hat{d}|2\rangle$ are the matrix elements of the dipole moment operator $\hat{d}$.

As is well known, in such a $\Lambda$ system near the energy splitting $(\omega_1-\omega_2)\approx\Delta$ a narrow two-photon dark resonance is observed. If the equality is exact, ($\omega_1-\omega_2$)=$\Delta$, a dark state exists:
\begin{equation}\label{dark2}
|dark\rangle=\frac{\Omega_2|1\rangle - \Omega_1|2\rangle}{\sqrt{|\Omega_1|^2+|\Omega_2|^2}}\,,
\end{equation}
in which the operator of atomic-field interaction $\widehat{V}_{}=-(\hat{d}E)$ equals zero.

We use the standard formalism of a Wigner density matrix $\hat{\rho}$(${\bf r},{\bf v}$), which satisfies the equation:
\begin{equation}\label{QME}
(\partial/\partial t+{\bf v}{\bf \nabla})\hat{\rho}+\widehat{\Gamma}\{\hat{\rho}\}=-(i/\hbar) [\widehat{V}_{},\hat{\rho}],\quad \text{Tr}\{\hat{\rho}\}=1,
\end{equation}
where ${\bf v}$ is the atom velocity, ${\bf \nabla}$ is the gradient operator. The functional operator $\widehat{\Gamma}\{\hat{\rho}\}$ describes the relaxation processes due to the spontaneous emission and the collisions (velocity changes, collisional broadening, spin-exchange) in a buffer gas or in a cell with an anti-relaxation coating.

Consider the scenario, when the spontaneous fluorescence intensity is fixed, i.e. the total population of atoms in the upper state $|3\rangle $ stays unchanged.
In this case, the condition (\ref{condition}) can be rewritten as an integral over the spatial and velocity variables:
\begin{equation}\label{spont}
S_\text{spont}\propto n_{at}(T)\int...\int f_T({\bf
v})\rho_{33}({\bf r},{\bf v})d^3{\bf r}d^3{\bf v}=const\,,
\end{equation}
where  $\rho_{33}({\bf r},{\bf v})=\langle 3|\hat{\rho}({\bf r},{\bf v})|3\rangle$, $f_T({\bf v})$ is a function of the Maxwell velocity distribution at temperature $T$, and  $n_{at}(T)$ is the spatial density of the atoms.

For simplicity, we study an optically thin medium with longitudinal size $L$, and neglect the Doppler effect. This corresponds to the case when the buffer gas collision broadening of the optical transitions $|1\rangle\leftrightarrow |3\rangle$ and $|2\rangle\leftrightarrow |3\rangle$ exceeds the Doppler broadening. Also, we consider a step-like cylindrical transverse field distribution:  $f(x,y)$=1 for $\sqrt{x^2+y^2}\leq r_0$ and  $f(x,y)$=0 for $\sqrt{x^2+y^2}> r_0$, where $r_0$ is the radius of the light beam. Under these assumptions the integral condition (\ref{spont}) can be reformulated for the single-atom population at the upper level:
\begin{equation}\label{spont2}
\rho_{33}=const\,.
\end{equation}
It is obviously that here we take into account the atoms, which are inside of the light beam.

The typical conditions for the two-photon spectroscopy of alkali atoms in the buffer gas are $\Gamma_\text{opt}\gg \gamma_\text{sp}\gg \Gamma_{0}$,  where $\Gamma_\text{opt}$ is the broadening of the optical transition, $\gamma_\text{sp}$ is the rate of spontaneous relaxation of the upper level $|3\rangle$ (see.  Fig.~\ref{Lambda}a), and $\Gamma_{0}$ is a rate of relaxation to the unperturbed equal distribution of populations at the ground states $|1\rangle$ and $|2\rangle$ (including the decay of the coherence $\rho^{}_{12}$=$\rho^*_{21}$ between states $|1\rangle$ and $|2\rangle$). For the optical resonance condition ($\delta_\text{1-ph}\ll \Gamma_\text{opt}$, see Fig.~\ref{Lambda}a) in case of the equal Rabi frequencies $\Omega_1=\Omega_2=\Omega$ and branching  decay rates $\gamma_{31}=\gamma_{32}=\gamma_\text{sp}/2$, the condition (\ref{spont2}) can be rewritten as:
\begin{eqnarray}\label{sp_alpha}
\rho_{33}&\approx &\frac{2\widetilde{\Omega}^2 [\widetilde{\Gamma}_\text{opt}(\widetilde{\Gamma}_{0}^2+ \widetilde{\delta}_\text{R}^2)+2\widetilde{\Gamma}_{0} \widetilde{\Omega}^2]}{\widetilde{\Gamma}_\text{opt}^2(\widetilde{\Gamma}_{0}^2 +\widetilde{\delta}_\text{R}^2) +2\widetilde{\Gamma}_\text{opt}(2\widetilde{\Gamma}_{0}+3\widetilde{\delta}_\text{R}^2)\widetilde{\Omega}^2+ 4\widetilde{\Omega}^4}\nonumber\\
&=&\alpha=const,\quad 0\leq\alpha <1\,.
\end{eqnarray}
The ``tilde'' (``$\sim$'')  implies that the variable has been normalized  by $\gamma_\text{sp}$ (i.e, $\widetilde{\Omega}\equiv\Omega/\gamma_\text{sp}$, etc.).

The expression (\ref{sp_alpha}) can be considered as an equation with respect to $\widetilde{\Omega}^2$, where the positive real root describes the frequency dependence of the squared Rabi frequency $\widetilde{\Omega}^2(\alpha, \delta_\text{R})$ on the two-photon detuning $\delta_\text{R} = (\omega_1-\omega_2-\Delta)$ for a given $\alpha$. Since the value $\widetilde{\Omega}^2$ is proportional to the field intensity $I$, we automatically obtain the frequency dependence of the input intensity $I(\alpha,\delta_\text{R})$ for a fixed number of excited atoms $N_e=\alpha n_{at}(T)\pi r^2_0 L$ in the light beam.

\begin{figure}[t]
\centerline{\scalebox{1.0}{\includegraphics{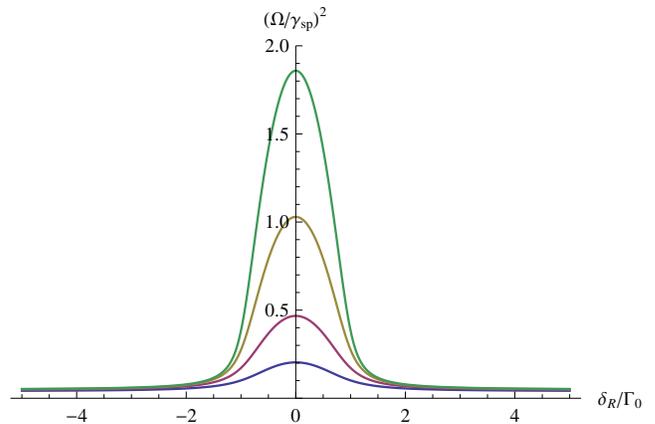}}}\caption{
The calculated $\widetilde{\Omega}^2(\alpha,\delta_\text{R})$  frequency dependence in case of  $\Omega_1 =\Omega_2$.
 Here have been used: $\delta_\text{1-ph}$=0, $\widetilde{\Gamma}_\text{opt}$=100, $\widetilde{\Gamma}_{0}$=0.001, $\gamma_{31}/\gamma_{32}$=1, $\alpha_{max}$$\approx$$0.001$, and   $\alpha$=0.0008,~0.0009,~0.00095,~0.00097 (from the bottom to the top).}\label{curves}
\end{figure}

\begin{figure}[t]
	\centerline{ \scalebox{0.9}{\includegraphics{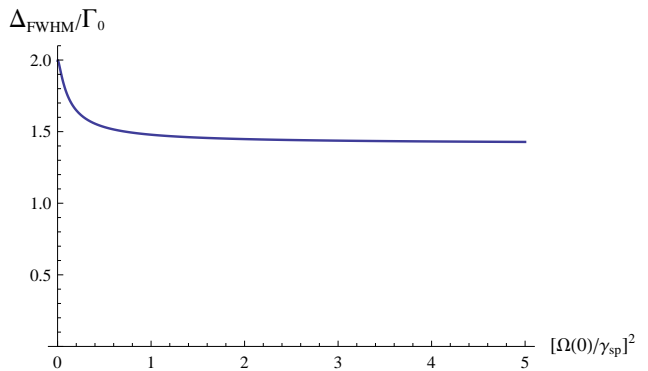} } }
	\caption{ The normalized CPT resonance width $\Delta_\text{FWHM}$ as a function of the  laser intensity $[\Omega(0)/\gamma_{sp}]^2$ at the two-photon resonance $\delta_\text{R}=0$ (the conditions are the same as in Fig.\ref{curves}). This demonstrates the possibility of the ``power narrowing'' for feedback resonance.}\label{broadening1}
\end{figure}

The calculated dependencies $\widetilde{\Omega}^2(\alpha,\delta_\text{R})$ for several values of the $\alpha$ parameter are presented in Fig.\ref{curves}. As follows from the plots, the resonance amplitude significantly increases for large $\alpha $ parameters. At the same time, far off the  resonance  the dependencies remain practically unchanged. The width of the resonance  is  less than $2\Gamma_0$ and is too remains almost unaltered. It can be shown that the resonance amplitude goes asymptotically to infinity. This extreme case takes place at
$\alpha\rightarrow\alpha_{max}$, where  $\alpha_{max}$ is the limit of $\rho_{33}$ at $\delta_\text{R}=0$ and
$\Omega^2\to\infty$. Thus, from (\ref{sp_alpha}) we find $\alpha_{max}\approx\widetilde{\Gamma}_{0}=\Gamma_{0}/\gamma_\text{sp}$. In general, $\alpha_{max}$ depends on the ratios $\Omega_1/\Omega_2$ and $\gamma_{31}/\gamma_{32}$, i.e. on the decay rates of the different channels (see Fig.~\ref{Lambda}a).
If $\alpha>\alpha_{max}$ there is an interval of
$\delta_\text{R}$ (centered at $\delta_\text{R}$=0), where the
positive real root $\widetilde{\Omega}^2(\alpha,\delta_\text{R})$
of equation (\ref{sp_alpha}) does not exist, i.e. the $I(\alpha,\delta_\text{R})$ has no physical meaning in this frequency interval.

The results, described above, could be explained in the following way. When $\alpha\approx\alpha_{max}\approx\widetilde{\Gamma}_{0}$ the input laser intensity is regulated by the feedback in a range between  $\Omega^2\gg\Gamma_\text{opt}\Gamma_{0}$ at the exact two-photon resonance ($\delta_\text{R}=0$) and  $\Omega^2 \approx\Gamma_\text{opt}\Gamma_{0}/2$, when the frequency is far off the  resonance ($|\delta_\text{R}|\to \infty$). This explains the significant increase of the resonance amplitude and contrast for $\alpha\to\alpha_{max}$.
The width of the resonance remains small in our method.
Indeed, let us consider the variation of the two-photon detuning from the tail to the center of the resonance. Obviously, under the week field conditions $\Omega^2 < \Gamma_\text{opt}\Gamma_{0}/2$ and at $|\delta_\text{R}|> \Gamma_{0}$ the population of the exited state $\rho_{33}(\delta_\text{R})$ and the intensity $I(\alpha,\delta_\text{R})\propto\widetilde{\Omega}^2(\alpha,\delta_\text{R})$ weakly depend on the frequency detuning.
Therefore, noticeable changes of the intensity $I(\alpha,\delta_\text{R})$ caused by detuning can be detected only at $|\delta_\text{R}|$ near $\Gamma_{0}$. In other words, the top of the resonance is formed at high light intensity and the bottom part at low. Thus, the quality of the resonance (the ratio of the amplitude and the width) also increases significantly. Additionally we note that the shape of the CPT resonance at $\alpha\rightarrow\alpha_{max}$ is far from a Lorentzian.

\begin{figure}[t]
\centerline{\scalebox{0.95}{\includegraphics{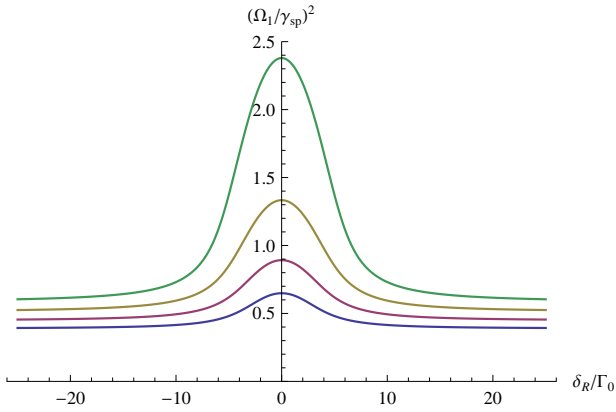}}}\caption{The calculated $\widetilde{\Omega}_1^2(\alpha,\delta_\text{R})$  frequency dependence in case of $\Omega_1 \neq\Omega_2$
 Here have been used:   $\delta_\text{1-ph}$=0, $\widetilde{\Gamma}_\text{opt}$=100, $\widetilde{\Gamma}_{0}$=0.001, $\Omega_1 /\Omega_2$=10, $\gamma_{31}/\gamma_{32}$=3, $\alpha_{max}\approx 0.00195$, and $\alpha$=0.0015,~0.0016,~0.0017,~0.0018 (from the bottom to the top). }\label{curves2}
\end{figure}

\begin{figure}[t]
\centerline{\scalebox{0.9}{\includegraphics{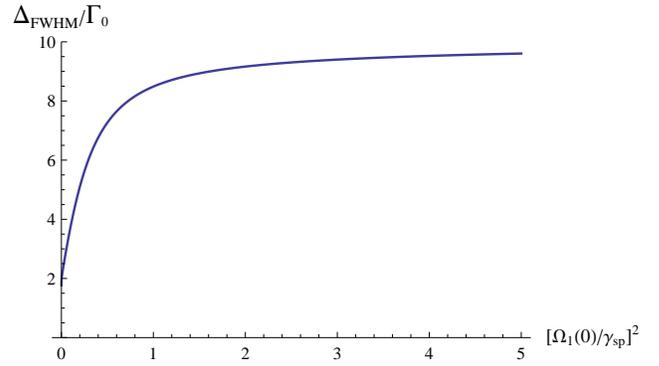}}}\caption{The normalized CPT resonance width $\Delta_\text{FWHM}$ as a function of laser intensity $[\Omega(0)/\gamma_{sp}]^2$ at the two-photon resonance $\delta_\text{R}=0$ (the conditions are the same as in  Fig.\ref{curves2}). }\label{broadening2}
\end{figure}

It should be pointed out that the behavior of the resonance width as a function of laser intensity is abnormal. In case of equal intensities of the bichromatic field components $\Omega_1=\Omega_2$ the resonance width \textit{narrows} in some range of the increasing intensity (see Fig.\ref{broadening1}). Such dependence is in contrast to the conventional spectroscopy, where the power broadening is regularly observed. Thus, the feedback spectroscopy allows, in principle, to overcome some fundamental limitations of the traditional spectroscopy (2$\Gamma_0$ minimal width of the CPT resonances in our case).
However, this ``extravagant'' result is not universal and corresponds to the
particular case of equal Rabi frequencies $\Omega_1=\Omega_2$. For
the case when $\Omega_1\neq\Omega_2$ the calculations show some
power broadening (see Fig.\ref{curves2}). But this broadening also has an abnormality - saturation at large intensities (see Fig.\ref{broadening2}).

In general, the analysis made above for feedback spectroscopy of the CPT resonance clearly shows a radical increase of the contrast and quality, which is not related to the optical density of the medium. Indeed, the frequency dependence of the input laser intensity $I(\alpha,\delta_\text{R})$ is independent of the number of atoms and can be extracted from the expression for the one-atom density matrix (see eq.~(\ref{sp_alpha})). On the other hand, in conventional spectroscopy the absorption is proportional to the number of atoms. Thus, the feedback-spectroscopy method is more effective (with respect to the conventional spectroscopy) for small atomic density. We believe that the number of atoms mostly influences on the noise properties of the feedback spectroscopy, but it requires a special theoretical and experimental study.

\begin{figure}[t]
\centerline{\scalebox{0.25}{\includegraphics{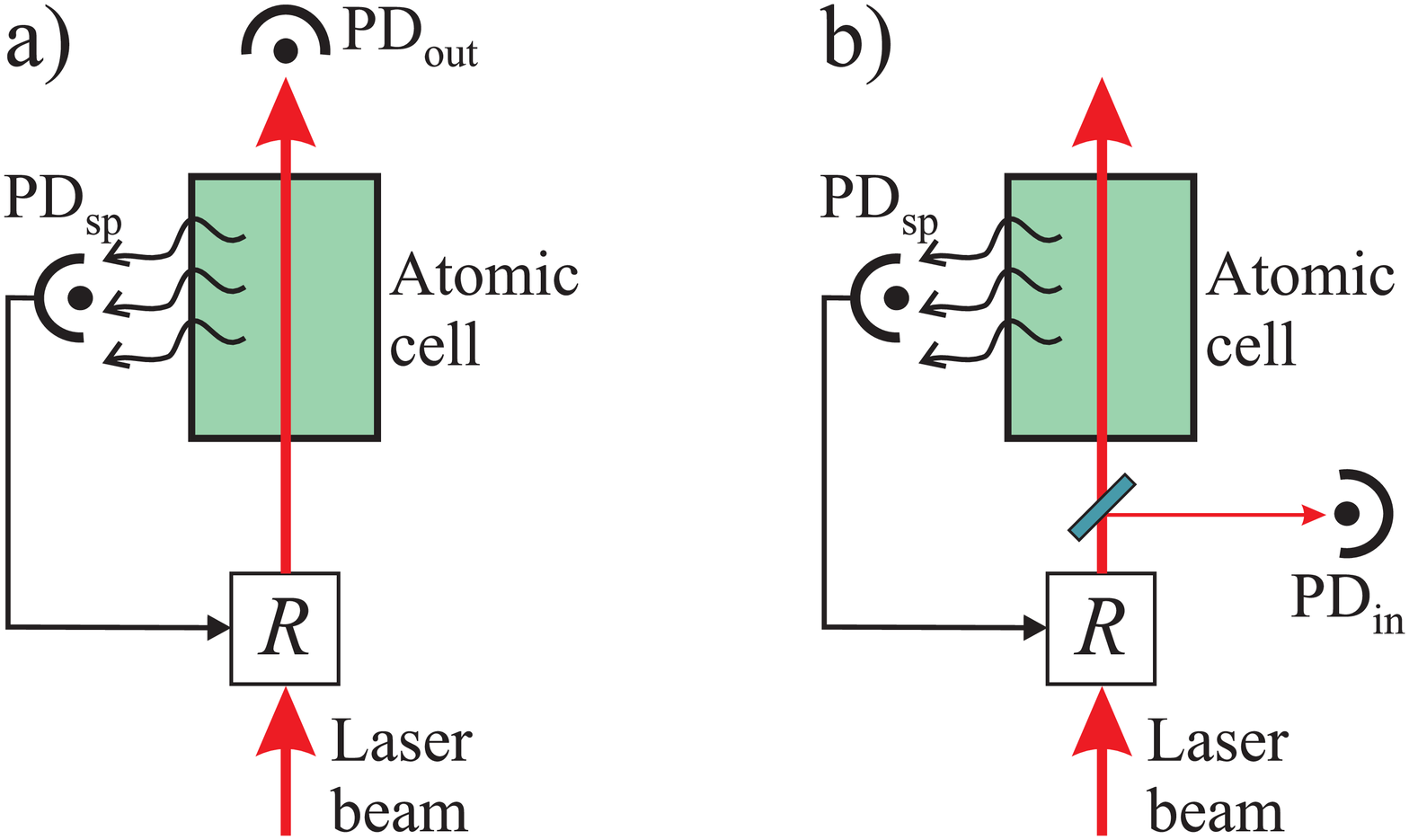}}}\caption{Detection of the atomic resonance in feedback spectroscopy:\\
a) PD$_\text{out}$ photodiode detects the transmitted light $S_\text{out}$;\\
b) PD$_\text{in}$ detects the input laser radiation $S_\text{in}$.\\
The  photodiode PD$_\text{sp}$ and device $R$ (e.g., acousto-optical modulator) form the feedback loop, sustaining the fluorescence at fixed level.}\label{two_schemes}
\end{figure}

This new spectroscopy method could be implemented in the following two ways: a) the transmission spectrum for the optically thin medium (see (Fig.$~\ref{two_schemes}$a) is  described as $S_\text{out}(\delta_\text{R})$$\propto$$[I(\alpha,\delta_\text{R})-B_{\alpha}]$, where the constant $B_{\alpha}>0$  corresponds to  the constant level of the spontaneously scattered light (for given experimental conditions); and b) the input laser intensity  $S_\text{in}(\delta_\text{R})$$\propto$$I(\alpha,\delta_\text{R})$ is detected directly before the cell (see Fig.~\ref{two_schemes}b).

\section{Experiment}

\begin{figure}[t]
\centerline{\scalebox{.5}{\includegraphics{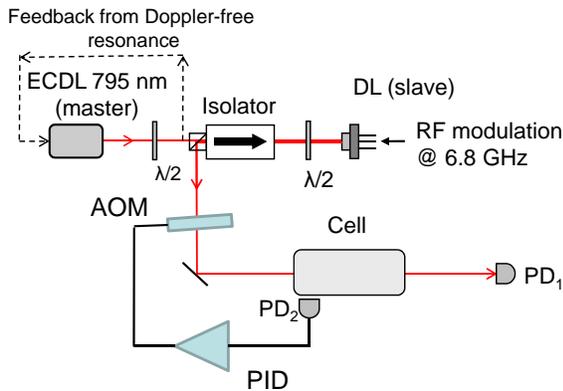}}}\caption{Setup.
 The array of the photodiodes PD2 detects the atomic vapor fluorescence. The feedback loop (that includes PID amplifier and acousto-optic modulator AOM) maintains the constant level
 of fluorescence at its value at the $\delta _\text{R}$=0.}\label{Setup}
\end{figure}

The experimental setup is shown in Fig.~\ref{Setup}. The high coherent resonant radiation of the extended cavity (ECDL) is injected into  the ``slave'' diode laser DL, which is modulated at the frequency of the hyperfine splitting $\Delta_{hfs}$=6.8~GHz for $^{87}$Rb (see. Fig.~\ref{Lambda}b).
The experiment is carried out with a Pyrex cell (25~mm long and 25~mm in diameter) containing isotopically enriched
$^{87}$Rb and a $\sim5$ Torr neon buffer gas. The cell is placed inside a magnetic shield. For the results reported here the cell temperature was in the range of 16--41~$^{\circ}$C.

The laser frequency is locked to the Doppler-free saturated absorption resonance by the DAVLL technique \cite{BudkerDAVLL}. The power of the laser radiation measured at the front window of the cell is in the 0.1--10~mW range, the beam diameter is 1--5 mm.
To excite the $\Lambda_{1,2}$-schemes in $Lin||Lin$ configuration (see. Fig.~\ref{Lambda}b) the carrier frequency is tuned to the $F_2 = 2\rightarrow F_e=1$ transition, and the high frequency side-band is tuned to the $F_1 = 1\rightarrow F_e=1$ transition \cite{lin_lin JETP}.

The ground state relaxation rate $\Gamma_0$ is inversely proportional to the transit time of atoms crossing the laser beam, enhanced by the multiple collisions with buffer gas molecules. Experimentally it has been found that this value is approximately $\Gamma_0\simeq$(0.5--1)~kHz. The broadening of the optical transition by the buffer gas is estimated as $\Gamma_\text{opt}\simeq$40--50~MHz \cite{Happer,Allard}.

\begin{figure}[t]
\centerline{\scalebox{1}{\includegraphics{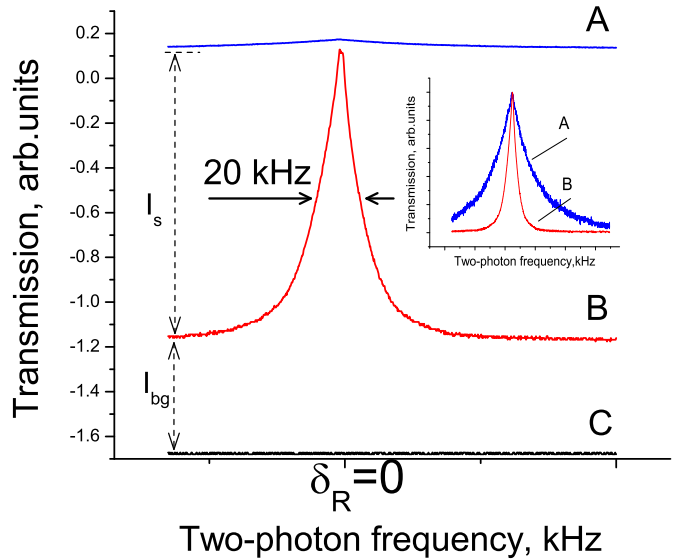}}}
\caption{CPT transmission for  unlocked ``A'' and  locked ``B'' the  fluorescence minimum intensity at $\delta_\text{R}=0$ frequency. The density of the atomic vapor is 3$\times$10$^{10}$~cm$^{-3}$ at the 24~$^{\circ}$C. The line ``$C$''  is the  no-light level.
The amplitude of the resonance ``$B$'' in $33$ times bigger than in case ``$A$''. The width of the resonance decreases from the 60~kHz (the case ``$A$'') to the 20~kHz (the case ``$B$''). Insert is the normalized transmission of the  resonance for both cases.}
\label{Three Resonances}
\end{figure}

The detector PD2 is  an array of 10 photodiodes connected in parallel and equally spaced along the circumference of the  the cell body. PD2 detects 1--5$\%$ of the fluorescence generated by the rubidium vapor. The PID amplifier sends the signal to the acousto-optical modulator (AOM) to lock the fluorescence intensity at the level of the $\delta _\text{R}$=0. The feedback loop has a unity gain up to $\sim30$~kHz. The detected EIT resonances are shown in Fig.$~\ref{Three Resonances}$, the curves correspond
to the $^{87}$Rb transmission spectra  for the two cases of ``A'' open and ``B'' closed fluorescence feedback loop with lock point at the maximum of the CPT transmittance. If the feedback loop is closed at frequencies different than $\nu=\Delta_{hfs}$ the
amplitude of the CPT resonance and the overall level of Doppler absorption increase (to be precise,  to the  level of non-resonant scattered light in the fluorescence signal). We detect an increase of the resonance amplitude and a decrease of the width, see Fig.$~\ref{Three Resonances}$. The contrast/width of the CPT resonance without and with a feedback loop are $2\%$/60~kHz (``A'') and $260\%$/20~kHz (``B'') correspondingly. Here we use the definition of the resonance contrast $C= I_\text{s}/I_\text{bg}$ as in \cite{Vanier-Review1}. The laser power and the beam sizes for this case are 6~mW and 5$\times$3~mm$^2$.

The scattered light from the window and the wall of the cell cause an additive shift of the fluorescence level. To avoid this distortion, the cell's surface are carefully cleaned. The magnitude of the scattered light  is 25 times less than the maximum of the fluorescence at the CPT resonance. The contrast (as well as the signal/noise ratio) increases with the gain of the feedback to the AOM.

\begin{figure}[t]
\centerline{\scalebox{1.2}{\includegraphics{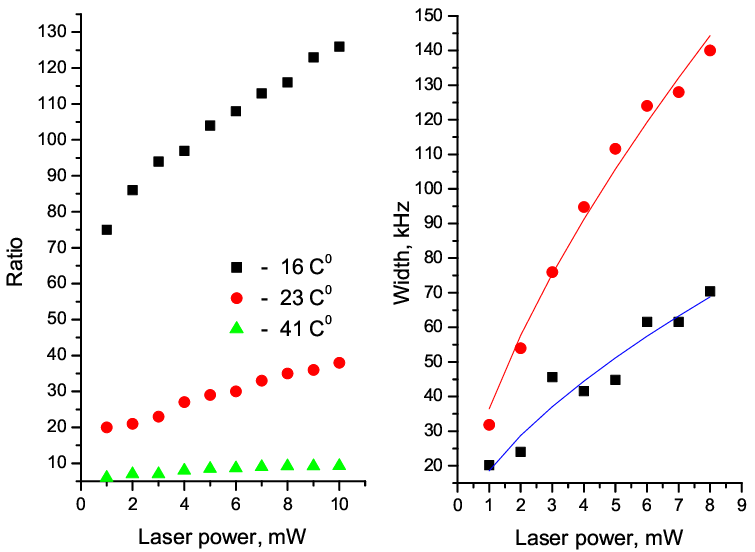}}}\caption{Left panel. The dependence of the  CPT resonances amplitude ratio  for locked and unlocked fluorescence  intensity  at the different temperatures $16 \,^{\circ}\mathrm{C}, 23 \,^{\circ}\mathrm{C}, 41\,^{\circ}\mathrm{C} $. \newline
Right panel. The power broadening of the CPT resonance for the conventional (top line) and feedback (bottom line) spectroscopy. Temperature of the atomic vapor is $23 \,^{\circ}\mathrm{C}$, beam size equals 1.5$\times$1.5 mm$^2$.} \label{Ratio}
\end{figure}

In Fig.$~\ref{Ratio}$ (left panel) the ratio of the EIT resonance amplitude for the cases of feedback and transmition spectroscopy are shown as a function of laser power for a set of different temperatures. One can see, that the lower atomic density profits from the feedback spectroscopy method. The CPT linewidth with the feedback is 1.5--3 times less than the one achieved by the means of conventional spectroscopy (right panel of Fig.$~\ref{Ratio}$). The intensities in the experiment range between 30--95~kHz or 0.36--4 in normalized $[\Omega(0)/\gamma_{sp}]^{2}$ units of Fig.$~\ref{broadening1}$. The fitting lines in the right panel have the weak nonlinearity probably due to the transversal Gaussian intensity distribution in accordance with the results of \cite{Taich_2004}.

In addition to the $Lin$$\parallel$$Lin$ field configuration \cite{lin_lin JETP}, the feeedback technique has been tested for other CPT schemes ($Lin$$\bot$$Lin$ \cite{Zanon} and standard $\sigma^+$$-$$\sigma^+$ configuration), as well as for the Hanle spectroscopy of the dark magneto-optical resonance \cite{Ron}. The saturation Doppler-free spectroscopy \cite{Demtroder} also has been tested. All of listed schemes show the same advantages of our method.

\section{Discussion and Conclusion}

In general, the performed experiments confirm the main results of our theoretical analysis. They show at low atomic vapor pressure a substantial increase of the resonance amplitude and a significant decrease of its width. However, instead of the ``power narrowing'' (Fig.~\ref{broadening1}) or saturation (Fig.~\ref{broadening2}) calculated at higher light intensities, our experiments demonstrate a typical power broadening, though at a slower rate compared with conventional spectroscopy methods. This contradiction with theory may come from our very simplified theoretical model. We have limited ourselves to a three-level system ignoring the rich hyperfine and Zeeman level structure of the $^{87}$Rb atom (see. Fig.~\ref{Lambda}b). Our calculations assume that the buffer gas broadening exceeds the Doppler broadening, but it is not the case of our experiments. Moreover, these calculations were made for the step-like transversal distribution of the light intensity instead of a spatially non-uniform profile (e.g, Gaussian). Additionally, the non-resonant light scattering off the cell windows can distort the feedback loop operation.

\begin{table}
\begin{tabular}{|c||ccc|}
\cline{1-3}
\hline
\multicolumn{4}{|c|}{Spectroscopy} \\
\hline
$\sharp$ & Feedback
|& Spontaneous |& Transmission\\
\hline
$C$ & 2.6 & 0.5& 0.02 \\
$q$ &$(2.6/\Gamma_{r})\times3$ &$ 0.5/\Gamma_{r}$ &$0.02/\Gamma_{r}$\\
$\sqrt{{I_\text{bg}~[\mu A]}}$ & 2.0  & 0.55 & 3.7\\
\hline
$\sigma$ &$\sigma_0$& $55\sigma_0$ &$200\sigma_0$\\
\hline
\end{tabular}
\caption{The contrast $C$, the figure of quality $q$=$C/\Gamma$ (here $\Gamma$ is the width of resonance) and the estimated instability $\sigma$ of  the CPT clock \cite{Vanier-Review2} for different detection scheme. The estimation has been  done for the shot noise limit and for the experimental conditions presented in Fig.~\ref{Three Resonances}, where $\Gamma_r$=60~kHz is the width of resonance without the feedback. The $\sigma_{0} $ is the clock instability locked to the CPT resonance given by the feedback spectroscopy technique.}
\label{T1}
\end{table}

To estimate the metrological perspectives of the feedback spectroscopy in such applications as atomic clocks and magnetometers, we refer to the formula for the frequency instability of the atomic clock $\sigma$ in the paper \cite{Vanier-Review2}. In this case, the shot noise instability limit could be expressed in terms of the resonance quality factor $q$ and the intensity of the light at the resonance wings $I_{bg}$: $\sigma\propto (I_{bg}\times q)^{-1}$. We have experimentally observed at identical conditions the CPT resonance in the three detection schemes - feedback and two conventional (transmission and spontaneous radiation) spectroscopies. The resulting contrast $C$, quality factor $q$, background current $I_{bg}$ and the {\em estimated} clock instability $\sigma$ are given in Table.~\ref{T1}. Using the feedback spectroscopy method the  $(\sqrt{I_\text{bg}}\times q)^{-1}$ product has been increased by two orders of magnitude compared to the traditional spectroscopy. This gives us a real hope for a significant improvement of the stability (sensitivity) of the CPT-based atomic clocks (magnetometers). Though, additional studies of the noise are required.

The data in Fig.~\ref{Ratio} (left panel) confirms another result of our analysis, that the feedback spectroscopy method is more profitable (with respect to the conventional spectroscopy) in less dense medium. Consequently, this gives the following advantages:\\
1). As the temperature is lowered the influence of the spin-exchange collision broadening reduces. So, at room temperature it is negligible smaller being by two orders than at 75~$^\circ$C, where the existing chip-scale atomic clocks operate \cite{Vanier-Review2}.\\
2). In hot cells the ongoing intrinsic reaction between the alkali metal and cell body (the ``curing'' process) poses a serious problem for the long term frequency stability of the clock \cite{Happer-Patton, Happer-Ma}. At lower temperatures the rate of this process is much smaller.\\
3). The Raman field parametric oscillations and other nonlinear effects are proportional to the number of atoms and are  reduced by two orders at room temperature if compared to their impact at 75~$^\circ$C \cite{Lukin}.

Also the following fundamental issue has caught our attention - can the electronic feedback alters the physical properties of the atomic medium in our method? For example, in the 1980s it was found that the application of  electronic feedback can change the essential physical property of a laser diode - its coherence \cite{Belenov,Yamamoto,Ohtsu}. If the feedback bandwidth is greater than the frequency spectrum of the dominate spontaneous noise, the linewidth of the diode laser is narrowed.  Similarly, we could expect to have an influence of the electronic feedback  on the physical properties of the medium due to the coupling between the spontaneous radiation and the laser light. In our case this coupling occurs because the fluctuations of the spontaneous emission (from all atoms) due to feedback are transferred into the laser light fluctuations and vice versa. Thus, from general viewpoint the spontaneous emission from the medium can not be considered as a simple sum from independent atoms. Consequently, under sufficiently wide-band feedback the fluctuation properties of the laser and spontaneous fields are correlated and should be considered matched. Therefore, the suggested  method needs further theoretical and experimental studies along these lines. Note, the theoretical treatment in the present paper did not account this correlation, which may be, in principle, an additional cause of disagreements between the experiment and the theory.

Finally, we have formulated the basic principles of the feedback spectroscopy.
This approach has been theoretically and experimentally verified on the CPT phenomenon, where a
robust increasing of the contrast and quality factor of the dark resonance was observed. In regard to CPT atomic clocks and magnetometers, the feedback spectroscopy might give great advantages. Different schemes, including magneto-optical Hanle spectroscopy, as well as  Doppler-free spectroscopy, also confirm the advantages of the suggested technique.

We thank A.V. Sivak and V.T. Miliakov for helpful discussions.
V.I.Yu. and A.V.T. were supported by RFBR (10-02-00591,
10-08-00844) and programs of RAS. D.I.S. and S.A.Z. were supported
by RFBR (12-02-31528).

V. I. Yudin e-mail address: viyudin@mail.ru

S. A. Zibrov e-mail address: serezha.zibrov@gmail.com

\end{document}